\documentclass[aps,prl,showpacs,twocolumn,superscriptaddress,a4paper]{revtex4}

\usepackage{graphics}

\begin{document}


\title{Formation of Nuclear ``Pasta'' in Supernovae
}


\author{Gentaro Watanabe}
\thanks{These two authors contributed equally.}
\affiliation{CNR INFM-BEC and Department of Physics, University of Trento, 38050 Povo, Italy}
\affiliation{RIKEN, 2-1 Hirosawa, Wako,
Saitama 351-0198, Japan}
\author{Hidetaka Sonoda}
\thanks{These two authors contributed equally.}
\affiliation{Department of Physics, University of Tokyo, Tokyo 113-0033, Japan}
\affiliation{RIKEN, 2-1 Hirosawa, Wako, Saitama 351-0198, Japan}
\author{Toshiki Maruyama}
\affiliation{ASRC, Japan Atomic Energy Agency, Tokai, Ibaraki 319-1195, Japan}
\author{Katsuhiko Sato}
\affiliation{Department of Physics, University of Tokyo, Tokyo 113-0033, Japan}
\affiliation{IPMU,
University of Tokyo, Kashiwa, Chiba, 277-8568 Japan}
\author{Kenji Yasuoka}
\affiliation{Department of Mechanical Engineering, Keio University, Yokohama, 223-8522, Japan}
\author{Toshikazu Ebisuzaki}
\affiliation{RIKEN, 2-1 Hirosawa, Wako,
Saitama 351-0198, Japan}



\date{\today}

\begin{abstract}
In supernova cores, nuclear ``pasta'' phases such as triangular
lattice of rod-like nuclei and layered structure of slab-like nuclei
are considered to exist. However, it is still
unclear whether or not they are actually formed in collapsing
supernova cores. Using {\it ab-initio} numerical simulations called
the Quantum Molecular Dynamics (QMD), we here solve this problem by
demonstrating that a lattice of rod-like nuclei is formed from a bcc
lattice by compression. We also find that, in the transition
process, the system undergoes zigzag configuration of elongated
nuclei, which are formed by a fusion of two original spherical nuclei.
\end{abstract}


\maketitle



The mechanism of collapse driven supernova explosions \cite{bethe} has been
a long-standing mystery \cite{colgate}.
Matter in supernova cores is also yet to be understood
and these issues are closely connected.
In the initial stage of the explosions, the collapsing iron core
experiences an adiabatic compression, which leads to an increase
of the central density from $\sim 10^{9}$ g cm$^{-3}$ to
around the normal nuclear density $\sim 3\times 10^{14}$ g cm$^{-3}$
(corresponding to the nucleon number density $\rho_{0}=0.165$
fm$^{-3}$) just before the star rebounds. 
Thereby, the Coulomb repulsion between
protons in nuclei, which tends to make a nucleus deform, becomes
comparable to the surface tension of the nuclei, 
which favors a spherical nucleus.  The pasta phases
(phases consisting of ``spaghetti''-like columnar nuclei and of
``lasagna''-like planar nuclei, etc.) \cite{bbp,rpw,hashimoto}
are thus expected to be formed
in the inner cores during the collapse of stars and amount to more
than 20\% of the total mass of the cores just before the bounce
\cite{opacity}.  Since the coherent scattering of neutrinos is very
different between pasta nuclei and spherical ones \cite{gentaro2,horowitz},
pasta phases can have a significant influence on the dynamics of
explosions \cite{horowitz,opacity}.  
Thus there is a growing interest
in pasta phases \cite{reviews}.

However, the above speculation that the pasta phases are formed in 
collapsing cores is based on phase diagrams of the equilibrium state
(e.g., Refs.\ \cite{bonche,lassaut,qmd_hot,newton,avancini} for non-zero temperatures)
or static and perturbative calculations \cite{review,iida}. 
Since formation of the pasta phases from a bcc lattice of spherical nuclei 
is accompanied by dynamical and drastic changes of the nuclear structure,
the fundamental question whether or not the pasta phases 
are formed in supernova cores is still open, and 
an {\it ab-initio} approach is called for.
In the present work, we answer this question using 
called the quantum molecular dynamics (QMD) 
\cite{aichelin,maruyama}.
QMD can properly incorporate the thermal fluctuations and is a
powerful and suitable approach \cite{note_shell}
to describe the deformation of nuclei 
in the present problem as has been exploited for studying 
the equilibrium phase diagram \cite{qmd,qmd_hot}
and the transition dynamics between the pasta phases \cite{qmd_transition}.

We use the QMD Hamiltonian of Ref.\ \cite{maruyama}
with the standard medium-equation-of-state parameter set 
\cite{note_hamiltonian}.
In our simulations, we consider a system with protons, neutrons, and
charge-neutralizing electrons in a cubic box with periodic boundary
condition.  The electrons are relativistic and degenerate, and 
can be well approximated as a uniform background 
\cite{review,screening}.
We calculate the Coulomb interaction by the Ewald sum.
Our simulations are carried out for 
the proton 
fraction $x\simeq 0.39$ and $0.49$ using different initial conditions
of the bcc lattice.  In the
following, we focus on the results for $x\simeq 0.39$ (the qualitative
results are the same for $x\simeq 0.49$).

The initial condition of $x\simeq 0.39$ is obtained in the following way.
We first prepare an isolated nucleus of $^{208}$Pb at zero temperature
and make a unit cell of the bcc lattice by setting two copies of the nuclei
in a box with the periodic boundary condition.
Using the Nos\'e-Hoover thermostat for
momentum-dependent potentials \cite{qmd_hot},
we then equilibrate this sample at the temperature $T=1$ MeV.
We combine eight replica of this bcc unit cell to make a
sample of the total number of nucleons $N=3328$
(with 1312 protons and 2016 neutrons) at the nucleon number density 
$\rho=0.15 \rho_0$ (the box size $L=51.23$ fm) and
equilibrate at $T=1$ MeV for $\simeq 7800$ fm$/c$ using the
Nos\'e-Hoover thermostat and further relax for $\simeq 4100$ fm$/c$
without the thermostat.
Equilibrating this sample at different
temperatures, we also prepare initial conditions for various
temperatures.


We simulate the compression of the bcc phase of spherical nuclei in
the collapse.  Starting from the above initial
conditions, we increase the density by changing the box size $L$ (the
particle positions are rescaled at the same time).  Here the average
rate of the compression is $\alt\mathcal{O}(10^{-6})\ \rho_0/($fm$/c)$ yielding
the time scale of $\agt 10^{5}$ fm$/c$ to reach the typical density
region of the phase with rod-like nuclei.
This is much larger than the time scale of the change of
nuclear shape (e.g., $\sim 1000$ fm$/c$ for nuclear fission) and thus the
dynamics observed in our simulation would be determined by the
intrinsic physical properties of the system, not by the density change
applied externally.
We perform adiabatic compression and isothermal compression 
at various temperatures \cite{note_simulations}.
In all the cases, we observe the formation of rod-like nuclei; 
here we show some typical examples in which we obtain a clear lattice 
structure of the rod-like nuclei.

\begin{figure}
\resizebox{8.2cm}{!}
{\includegraphics{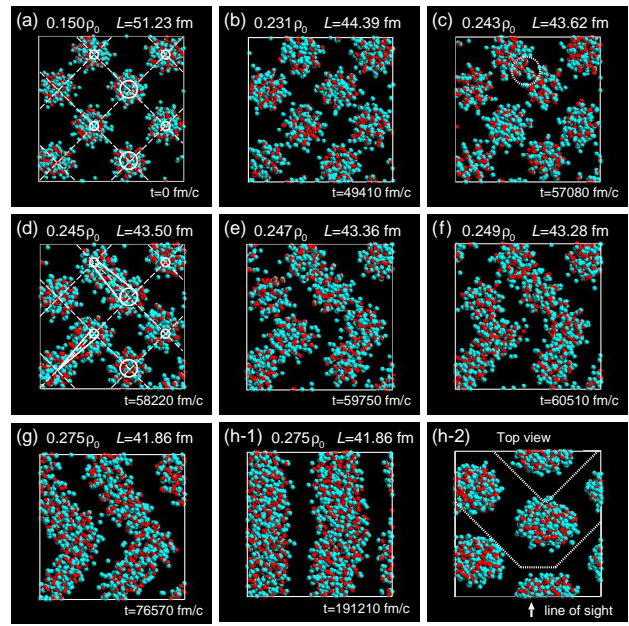}}
\caption{\label{snapshot}(Color online) Snapshots (two-dimensional
projection) of the transition process from the bcc lattice of
spherical nuclei to the pasta phase with rod-like nuclei.  The red
particles show protons and the green ones neutrons.  In panels (a)-(g)
and (h-1), nucleons in a limited region 
relevant for the two rod-like
nuclei in the final state (h) [surrounded by the dotted lines in panel (h-2)]
corresponding to eight nuclei in the initial condition (a)
are shown for visibility.
The vertices of the
dashed lines in panels (a) and (d) show the equilibrium positions of
nuclei in the bcc lattice and their positions in the direction of the
line of sight are indicated by the size of the circles: 
vertices with a large circle, with a small circle, and those without a
circle are in the first, second, and third lattice plane,
respectively.  The dotted circle in panel (c) show the first pair of
nuclei start to touch.  The solid lines in panel (d) represent the
direction of the two elongated nuclei: they take zigzag configuration.  
In the final state [(h-1)
and (h-2)], almost perfect triangular lattice of rod-like nuclei is
obtained.
The box sizes are rescaled to be equal
in the figures.
}
\end{figure}

Figure \ref{snapshot} shows the snapshots of the formation process of
the pasta phase in adiabatic compression.  
Here, we start from the initial condition of 
$x=0.39$ and $T=0.25$ MeV ($t=0$ fm$/c$).
At $t\simeq 57080$
fm$/c$ and $\rho\simeq 0.243 \rho_0$ [Fig.\ \ref{snapshot}(c)], 
the first pair of
two nearest-neighbor nuclei start to touch and fuse (dotted circle), 
and then form an elongated nucleus [see, e.g., Fig.\ \ref{snapshot}(d)].  
After multiple pairs of nuclei become such elongated nuclei, 
we observe zigzag structure as shown in Fig.\ \ref{snapshot}(d).
These elongated nuclei start to stick together at
$t\simeq 59000$ fm$/c$ and $\rho=0.246 \rho_0$, and all the 
nuclei fuse to form rod-like nuclei at $t\alt 72700$ fm$/c$ and
$\rho\alt 0.267 \rho_0$.
At $t= 76570$ fm$/c$ and $\rho= 0.275 \rho_0$, we stop the compression
($T\simeq 0.5$ MeV in this stage).
We then relax the system microcanonically for $65510$ fm$/c$ 
and further relax at $T=1$ MeV for $\simeq 10000$ fm$/c$ 
(after that we take back the temperature to $\simeq 0.5$ MeV 
and relax for $\simeq 20000$ fm$/c$; 
for the increase and decrease of the temperature between $0.5$ and 1 MeV, 
we take $\simeq 5000$ and $\simeq 15000$ fm$/c$, respectively).
Finally, we obtain a triangular lattice of
rod-like nuclei [Figs.\ \ref{snapshot}(h-1) and (h-2)]. 
Remarkable point is that, in the middle of the transition process, 
pair of spherical nuclei get closer to fuse 
in a way such that the resulting elongated nuclei 
take a zigzag configuration (hereafter, we call it 
``zigzag pairing'') and they further
connect to form wavy rod-like nuclei.
This feature is observed in all the other cases in which we obtain a
clear lattice structure of rod-like nuclei (among them, 
the highest temperature case is the adiabatic compression from $T=1.5$
MeV), and the above scenario of the transition process is
qualitatively the same also for those cases.  It is very different
from a generally accepted picture 
(see, e.g., p.~462 of Ref.\ \cite{review}) that all the nuclei
elongate in the same direction along the global axis of the resulting
rod-like nuclei and they join up to form straight rod-like nuclei.

It has been regarded so far that the formation of the pasta phases in
supernova cores is triggered by the fission instability with respect
to the quadrupolar deformation of spherical nuclei \cite{review}.
To examine this point, we investigate the variance of the radius of
each nuclei over the solid angle and the area-averaged mean curvature
of the nuclear surface; 
neither of these quantities show
a significant increase throughout the compression process until the
nuclei start to touch [see also Fig.\ \ref{snapshot}(b)].  
This means that, before nuclei deform to be
elongated due to the fission instability, they stick together keeping
their spherical shape.
This is consistent with the result of Ref.\ \cite{brandt}
(see also Ref.\ \cite{burvenich}),
which shows that, within an incompressible liquid-drop model, the
condition for the fission instability is not satisfied if one takes
account of the background electrons.

\begin{figure}
\rotatebox{0}{
\resizebox{!}{6.6cm}
{\includegraphics{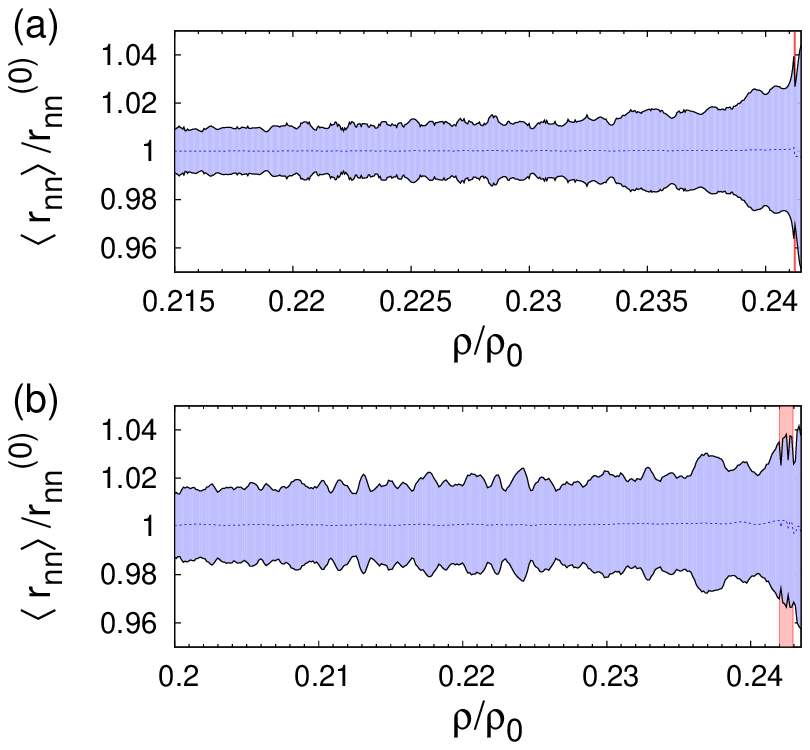}}}
\caption{\label{avrdis}(Color online) 
Distance $r_{\rm nn}$ between centers of mass of nearest-neighbor nuclei
in the compression process.
The blue line shows the mean value $\langle r_{\rm nn} \rangle$
and the blue filled region shows the standard deviation $\Delta r_{\rm nn}$ 
of $r_{\rm nn}$ normalized by $r_{\rm nn}^{(0)}$ 
(a value of $r_{\rm nn}$ for a perfect bcc lattice 
without displacement of nuclei).
Panel (a) is for the isothermal compression at $T=0$ MeV 
and panel (b) is for the adiabatic compression starting from 
$T=0.25$ MeV.
The first pair of nuclei touch at a density in the red filled region
(corresponding to $r_{\rm nn}^{(0)} \simeq 18.9$ fm for both cases)
whose width shows its uncertainty.
}
\end{figure}

To understand the mechanism which triggers the formation of the pasta phases,
we calculate the distance $r_{\rm nn}$ 
between centers of mass (for protons) of nearest-neighbor nuclei
in the compression process of our simulations \cite{note_cluster}.
Figure \ref{avrdis} shows the mean value $\langle r_{\rm nn} \rangle$
and the standard deviation $\Delta r_{\rm nn}$ of 
$r_{\rm nn}$ calculated for 16 nuclei in the simulation box. 
Let us first focus on the result obtained for the isothermal compression 
at $T=0$ MeV 
[Fig.\ \ref{avrdis}(a)], 
where we can see a clear signature due to the absence of thermal 
fluctuations.  We note that there is a significant increase of 
$\Delta r_{\rm nn}$ just before the nuclei start to connect. 
This shows that, before pairs of nuclei touch, nuclei displace from 
the equilibrium position of the bcc lattice and the bcc structure is 
spontaneously broken.  As a result, each pair of nuclei approach 
to fuse and then form an elongated nuclei arranged in a zigzag configuration. 
At non-zero temperatures, thermal fluctuation smears the above signature
and it would also assist triggering the formation of the pasta phases;
however even in the case of the adiabatic compression starting at $T=0.25$ MeV,
we can still see a significant increase of $\Delta r_{\rm nn}$
just before the nuclei touch and fuse [Fig.\ \ref{avrdis}(b)].

\begin{figure}
\resizebox{7.7cm}{!}
{\includegraphics{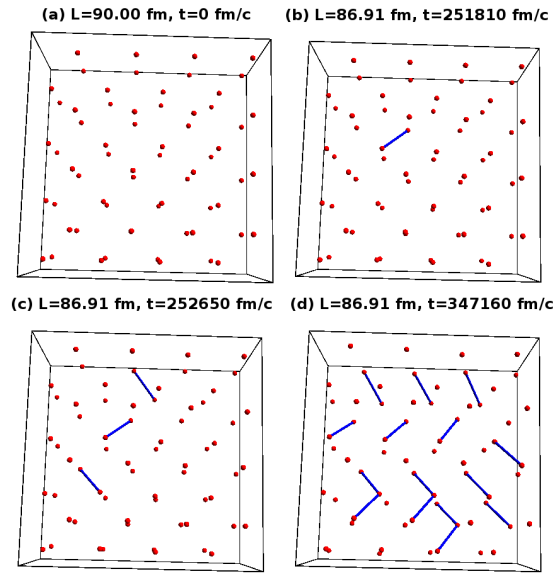}}
\caption{\label{simplemodel}(Color online) Snapshots of the transition process 
in the simulation of isothermal compression at $T < 0.01$ MeV 
using the simplified model.
Starting from the bcc lattice of nuclei (a),
the first pairing (b),
zigzag pairing around the first pair (c),
and finally a zigzag structure (d) are observed.
The red particles show the centers of mass of nuclei and 
the nuclei within the distance less than $0.89 r_{\mathrm{nn}}^{(0)}$
are connected by a blue line. 
Nuclei and connections within only two lattice planes 
normal to the line of sight are shown.
}
\end{figure}

Using a simplified model, we now examine our results of the QMD 
simulations. 
When nearest neighbor nuclei are so close 
that the tails of their density profile overlap with each other,
net attractive interaction between these nuclei starts to act 
due to the interaction between nucleons in different nuclei 
in the overlapping surface region.  We consider that this 
attraction between nuclei leads to the spontaneous
breaking of the bcc structure.
In order to examine this hypothesis, we consider a minimal model 
in which each nucleus is treated as a point charged particle 
interacting through the Coulomb potential and the
potential of the Woods-Saxon form: 
$V(r)=V_0\left\{1+ \exp\left[(r-R)/a\right]\right\}^{-1}$,
which describes the finite size of nuclei 
and models the 
internuclear 
attraction when the nearest neighbor nuclei start to touch.
Since nuclei start to connect before they are 
deformed, it is reasonable to treat a 
nucleus as a sphere and incorporate only its center-of-mass degree of 
freedom.
Parameters $R$ and $a$ represent the interaction radius of a nucleus and range
of the effective nuclear forces, respectively and we take
$V_0= -11.5$ MeV, $R=17$ fm, and $a=0.5$ fm \cite{note_WSpot}.
With this model, we carry out the isothermal 
(keeping the system cool at $T < 0.01$ MeV by the frictional relaxation method
throughout the simulation; this small but non-zero temperature is sufficient 
for symmetry breaking)
compression of the bcc lattice of 128 nuclei of
$^{208}{\rm Pb}$, which corresponds to 
8 times larger system than that of the QMD simulations.
In Fig.~\ref{simplemodel} we show the snapshots of this simulation.
Here we compress from $0.221 \rho_0$ ($L=90.00$ fm)
at a rate of $\dot{\rho}<\mathcal{O}(10^{-6})\ \rho_0/(\mathrm{fm}/c)$.
At $0.246 \rho_0$ ($L=86.91$ fm)
the first pair of nuclei starts to get closer [Fig.~\ref{simplemodel}(b)]
and then we stop the compression and relax the system. 
We observe zigzag pairing around the first pair [Fig.~\ref{simplemodel}(c)]
and finally we obtain a zigzag structure [Fig.~\ref{simplemodel}(d)].
This supports that the zigzag structures observed in the QMD simulations 
are not caused by a finite size effect of the simulation box.

In conclusion, we have shown that an ordered structure of rod-like nuclei
can be formed by compressing a bcc lattice of spherical nuclei.  
Our result establishes that the pasta phases can be formed in 
collapsing supernova cores.  Unlike a generally accepted conjecture, 
we have observed that
nuclei start to connect before they deform due to the fission instability.
This spontaneous breaking of the bcc structure is due to an 
attraction between nuclei caused by the overlap of the tails of 
nucleon distribution of neighboring nuclei.
We have also discovered that, in the transition process, 
the system takes a zigzag configuration of elongated nuclei, 
which are formed by a fusion of original two spherical nuclei.
Since a drastic change of the neutrino transport in the pasta phases 
pointed out in Ref.\ \cite{gentaro2} has been already shown
\cite{horowitz} (see also Ref.\ \cite{opacity}), 
it is very interesting to perform
core collapse simulations incorporating this effect as the next step.

\begin{acknowledgments}
We are grateful to Chris Pethick and K. Iida for helpful discussions
and comments.  
We used MDGRAPE-2 and -3 of the RIKEN Super Combined Cluster System.
This work was supported in part by the JSPS and
by the MEXT through Research Grants No. 19104006.
\end{acknowledgments}


\begin{thebibliography}{99}
\bibitem{note_shell} 
Shell effects, which are not incorporated in QMD, are
less important 
since they are washed out by thermal fluctuations at several MeV 
at subnuclear densities \cite{newton}.
\bibitem{note_simulations}
For $x=0.39$, we perform isothermal compression 
at $T=0$, 0.1, and 0.25 MeV, and adiabatic compression starting from 
$T=0.25$, 0.5, 0.75, and 1 MeV. (For $x=0.49$, we carry out
adiabatic compression starting from $T=0.25$, 0.5, 0.75, 1, and 1.5 MeV.)
In the isothermal compression at $T=0$ MeV, we keep the system cool
by frictional relaxation method (see, e.g., Refs.\ \cite{maruyama,qmd}) 
throughout the compression process (as a result, $T<0.01-0.04$ MeV).
In the isothermal compression at non-zero temperatures, we keep the 
temperature using the Nos\'e-Hoover thermostat.
(In the adiabatic compression, we compress the system without any thermostat.)
\bibitem{note_cluster}
Here, we employ the clustering algorithm:
we identify protons which are within the threshold
distance $r_{\rm th}$ from at least one of them are in the same
nucleus.
We take $r_{\rm th}=3.5$--$4$ fm and have checked that
the results do not change by varying $r_{\rm th}$ around this value.
In calculating $\langle r_{\rm nn} \rangle$ and $\Delta r_{\rm nn}$,
nuclei which connect to the other ones are excluded.  The sudden
reduction of $\Delta r_{\rm nn}$ when the first pair of nuclei touch
is due to this treatment.
\end{thebibliography}
\end{document}